\documentstyle[editedvolume,numreferences,graphicx,floatflt]{crckapb}
\newcommand{\stt}{\small\tt}
\begin{opening}

\title{Telescope Bibliometrics 101}
\author{Uta Grothkopf}
\institute{European Southern Observatory\\
           Karl-Schwarzschild-Stra{\ss}e 2\\
           D-85748 Garching, Germany\\
           \stt esolib@eso.org}
\author{Jill Lagerstrom }
\institute{Space Telescope Science Institute\\
           3700 San Martin Drive\\
           Baltimore MD 21218, USA\\
           \stt lagerstrom@stsci.edu}

\end{opening}

\begin{document}
\setcounter{page}{1}

\begin{abstract}
During recent years, bibliometric studies have become increasingly important in evaluating individual 
scientists, institutes, and entire observatories. In astronomy, often librarians are involved in maintaining 
publication databases and compiling statistics for their institutions. In this paper, we present a look 
behind the scenes to understand who is interested in bibliometric statistics, which methodologies astronomy 
librarians apply, and what kind of features next-generation bibliographies may include.
\end{abstract}

\section{Introduction}

There are many ways to assess research output, for instance by investigating how many research grants have 
been received, when and where research has been presented at conferences, or how many students graduated 
under the supervision of specific researchers. Another common tool is bibliometric studies, i.e., using 
metrics to measure productivity and impact through publications and citations. 

Bibliometric studies have quite some history; studies go at least back to the 1960s when the Science 
Citation Index was first issued. A large number of papers have been published on this topic in general 
and more specifically in the context of astronomy.

In this paper, we will focus on the following aspects:
\begin{itemize}
\item
Bibliometric studies -– what are they and who is interested?
\item
Linking publications and data -– how does this happen, and where can interested users get access?
\item 
Telescope bibliographies –- who compiles them, and how? What are the current tools and methodologies? 
\end{itemize}

Finally, we will take a look ahead to see what might be coming next.

\section{Typical bibliometric measures}

Many bibliometric studies use one or several of the following measures. However, they all have some 
advantages as well as some disadvantages.
\begin{itemize}
\item
\underbar{Number of publications}: measures productivity, but does not report anything about the impact
\item
\underbar{Number of citations}: gives information about the impact, but can be inflated due to many 
reasons, for instance incorrect or incomplete citations, as well as biased citing behavior (authors 
citing well-known authors rather than young, unknown ones)
\item
\underbar{Mean or median cites per paper}: this measure is better suited for comparisons of scientists 
or facilities that have been active for different numbers of years, but it seems to reward low productivity
\item
\underbar{`High-impact papers'}: this metric has been introduced in astronomy by Juan P.~Madrid, then at the 
STScI, a few years ago. Basically, Madrid used the ADS to retrieve information about the 200 most-cited 
papers in a given publication year, identified those that were based on observational data and calculated 
the impact of facilities (telescopes, observatories) of each paper. The results per paper were added, and 
the facilities ranked by contribution to this set of {\em Top 200} papers. The drawback of this method is 
that so-called hot topics are favored and can outnumber all other facilities for instance in a year of data 
release. The method is also time-consuming and, to a certain extent, subjective as the contribution 
percentage is assigned by the bibliometric researcher
\item
\underbar{{\em h}-index}: the {\em h}-index is meant to combine metrics for productivity and impact 
(Hirsch 2005); {\em h} itself is not suited for comparisons as it does not contain information about 
the number of years of operation. For comparisons another value, the so-called {\em m} parameter, also 
described by Hirsch, should be used. 
\end{itemize}

All measures have to be applied with greatest care and never in an isolated way as they only shed light 
on a very limited area of performance of research output. If used for comparisons, several metrics should 
be applied in parallel to get a more complete picture. 

\section{Users of bibliometric studies}

A wide range of groups are interested in bibliometrics, including instrument scientists, management of 
observatories, governing bodies and funding agencies who want to 
\begin{itemize}
\item
evaluate the performance of telescopes and instruments
\item 
measure the scientific output from observing programs
\item
define guidelines for future facilities
\item
compare in-house facilities with other observatories and telescopes
\item
interconnect resources, for instance in the context of Virtual Observatory projects
\end{itemize}

Establishing telescope bibliographies is the prerequisite to bibliometric studies in astronomy; it closes 
the loop from (a) observing programs carried out by astronomers, (b) data stored in an archive, (c) 
scientific papers which use the data, and (d) records in telescope bibliography databases that connect 
the papers with the data through observing programs (program IDs). Archives can be searched by program ID, 
and all papers published so far that use the data these programs generated can be listed. This procedure 
assures a maximum return of science benefits from observing proposals, for instance by making archived data 
easily available to other interested researchers. 

Telescope bibliographies can be accessed in various ways. Firstly, a listing has been compiled by the ESO 
Library and is available on the web\footnote{\tt http://www.eso.org/libraries/publicationlists.html}; 
it provides links to the databases of many large observatories. The user interfaces of telescope 
bibliographies of all major observatories provide the option to search by observation or data set number. 
If these numbers are noted in the bibliographies' records, a listing of all published articles using the 
data is just a click away. 

For example, in the case of ESO, the public interface allows to search for specific programs, or for papers 
that use data from certain instruments. The result will be a list of papers for which the facilities that 
were used are shown, as well as the program IDs. A click on the program ID will take users to the observing 
schedule from where more information on the proposals can be accessed. The data can be requested, or other 
papers using the same program can be found. 

Another way of finding lists of papers that use observational data from specific observatories is to use 
the ADS {\em Filters} section which can be found in the lower part of the main ADS search 
screen\footnote{\tt http://adsabs.harvard.edu/abstract\_service.html}. The option 
{\em ``All of the following groups"} as well as at least one of the listed facilities need to be 
selected. The records contained in the results set will all have {\em D} (for data) links which link 
back to the underlying observations.

\section{Compiling telescope bibliographies}

In order to compile such literature lists with links to data, or telescope bibliographies, certain 
prerequisites are necessary (Fig. 1).

\begin{figure}
\centerline{\includegraphics[height=10cm,angle=0]{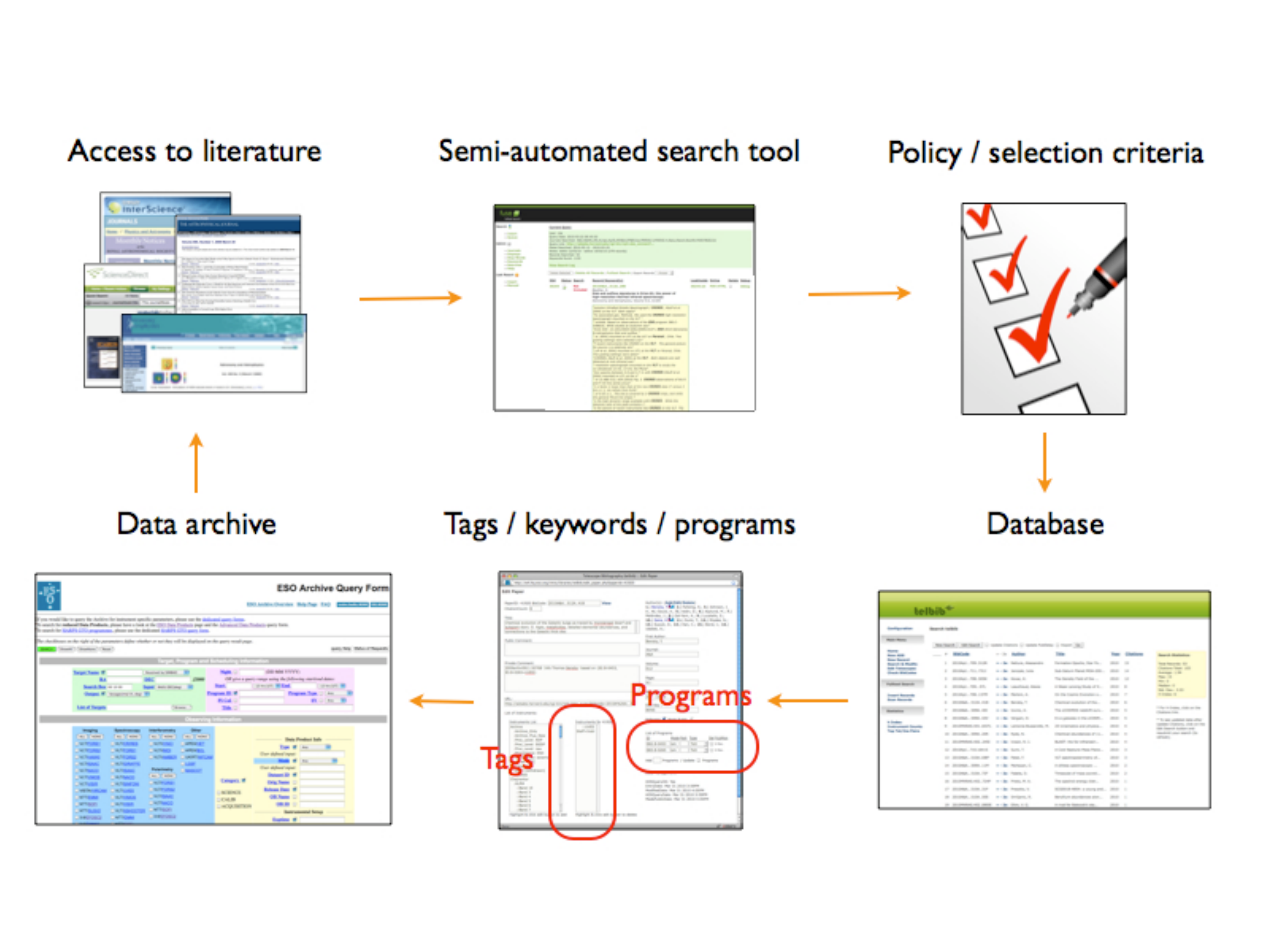}}
\caption{Telescope bibliographies close the loop from published literature to data located in archives, 
and back.}
\end{figure}

Compilers need to have access to the literature. This can be in the form of print versions of scientific 
journals, or electronic access. The latter can be established by either pulling over PDFs to screen them 
locally, or by searching full-texts at the publishers' sites. Ideally, a semi-automated search tool is 
available that helps compilers to pinpoint relevant articles. This will be explained in more detail below. 
If such a search tool is not in place, PDFs have to be searched one-by-one, or the paper version has to be 
inspected visually. 

Most importantly, a policy needs to be in place that determines which papers shall be included in the 
telescope bibliography, and which are to be excluded. The topic of selection criteria can be handled quite 
differently among large observatories. Once relevant articles are identified, a database has to be 
established to host records with bibliographic information about the published papers. 

Observatories will have their own individual set of tags and keywords which are assigned to records to 
describe the content of the papers as well as the facilities that generated the data. Specific information, 
such as program IDs, will be added at this stage, too. These identifiers link to the observatories' archive, 
and from there back to the published articles.

\section{Staffing situation, journals screened, and search strategies}

An additional very important ``ingredient" needed to establish and maintain telescope bibliographies is 
staff. In many observatories, librarians are involved in the process, or are even the main person in charge. 

In early 2010, Jill Lagerstrom, librarian at STScI, conducted a survey among 16 large observatories 
(Lagerstrom 2010). A total of 14 institutions replied, namely CFHT, Chandra, FUSE/Galex/IUE, ESO, Gemini, 
HST/STScI, Isaac Newton Group, Keck, NOAO, NRAO, SDSS, Spitzer, Subaru, and XMM Newton. The survey revealed 
that large observatories involve between one and four staff members to compile their bibliographies, even 
though not all of them full-time. 

The survey also focused on the journals which are searched by these institutions. Respondents provided a 
combined list of more than 30 journals which are screened, but surprisingly, only four are searched by all 
survey participants: {\em A\&A}, {\em AJ}, {\em ApJ} and {\em Supplement}, as well as {\em MNRAS}. 

The full-text can be accessed using one or several of the following methods:
\begin{itemize}
\item
\underbar{Print journals}: screening the full-text by using the print version of journals provides access 
to the entire text of papers including footnotes, figures, captions, etc. However, it depends on the skills 
and capabilities of the compilers to detect all instances of relevant keywords and facilities in the text to 
gather a complete list of published articles
\item
\underbar{E-journals (PDFs)}: this methodology implies retrieving the PDF format of articles from the 
publishers' web sites and searching them locally. Like with method 1, all sections of papers are screened.
\item
\underbar{Journal web sites}: many publishers provide an option to search full-texts of articles directly at 
their web site. While this functionality can indeed be very useful, it is usually not immediately 
understandable for users whether this feature actually searches the entire text, or whether certain sections 
are not indexed and can therefore not be searched properly. 
\item
\underbar{ADS abstracts}: at present (May 2010), ADS provides full-text searches only for historic 
literature; access to recent publications includes only title, abstract, and selected footnotes. 
If compilers of telescope bibliographies rely on ADS abstract searches for their work, they must be 
aware that almost inevitably they will miss many relevant papers. 
\item
\underbar{Author self-reporting}: some compilers use reports from authors as an additional or even exclusive 
means to find papers. It must be noted that, if applied as the sole method for collecting articles, relying 
on author self-reports will reveal only a fraction of all relevant papers; it is therefore the least 
recommended way of compiling telescope bibliographies. 
\end{itemize}

The survey also investigated whether compilers of bibliographies use similar search statements, and how 
elaborate and comprehensive they are in their attempts to identify papers. Not surprisingly, all respondents 
stated that they search for their observatory's telescopes, typically both the abbreviation as well as the 
full name. Some are more specific and include instrument names, as well as surveys, archives, or specific 
science programs (e.g., GOODS). For some institutions, it can be useful to search for geographic locations 
(Green Bank, Paranal, etc.) or even for concepts (e.g., X-ray). 
 
The final survey question aimed at the biggest challenge compilers face in gathering papers. The majority 
of respondents stated that their biggest problem is authors who do not provide sufficient details about the 
source of their data, so that the programs that generated the data cannot be traced without doubt. Others 
replied that they face problems with restrictive publishers when they access large numbers of articles at 
their web sites. Many compilers feel that the amount of work time they can dedicate to their respective 
telescope bibliography is not sufficient in order to complete an exhaustive search. The idea of having 
a central search option for full-texts at the ADS was ventilated; this would be more time and 
resources-efficient than duplicating retrieval efforts locally at each observatory, as is currently done.  

It became evident that the most difficult part is to define a policy to clearly govern which papers fulfill 
the criteria for inclusion, and which do not, as well as consistency in applying this policy. Recently, a 
group has been formed, initiated by the compilers of telescope bibliographies at the Chandra Archive, 
STScI, and ESO, to provide best practices for maintaining bibliographies, to develop recommendations for 
using cross-facility bibliometrics, and to share solutions to common 
problems\footnote{\tt http://groups.google.com/group/astrobib/}. 

\section{Tools of the trade: FUSE and telbib}

While compilers of bibliographies in the past had to rely on their own abilities to retrieve relevant 
papers by visually scanning the literature, lately some tools have become available to support them in 
their work. One of these tools is the ESO Full-Text Search tool (FUSE) which has been developed by the 
ESO Library. Based on an ADS search, FUSE pulls over PDFs from publishers, converts them into text, and 
searches for keywords and text strings chosen by the individual observatory. If keywords are detected, 
they are shown in context (two lines of text) for inspection by the compilers. The excerpts often reveal 
immediately whether the highlighted paper should be investigated in more detail, or whether the keywords 
are used in a context that is not relevant for the compilation of the telescope bibliography (Fig. 2). 

\begin{figure}
\centerline{\includegraphics[height=10cm,angle=0]{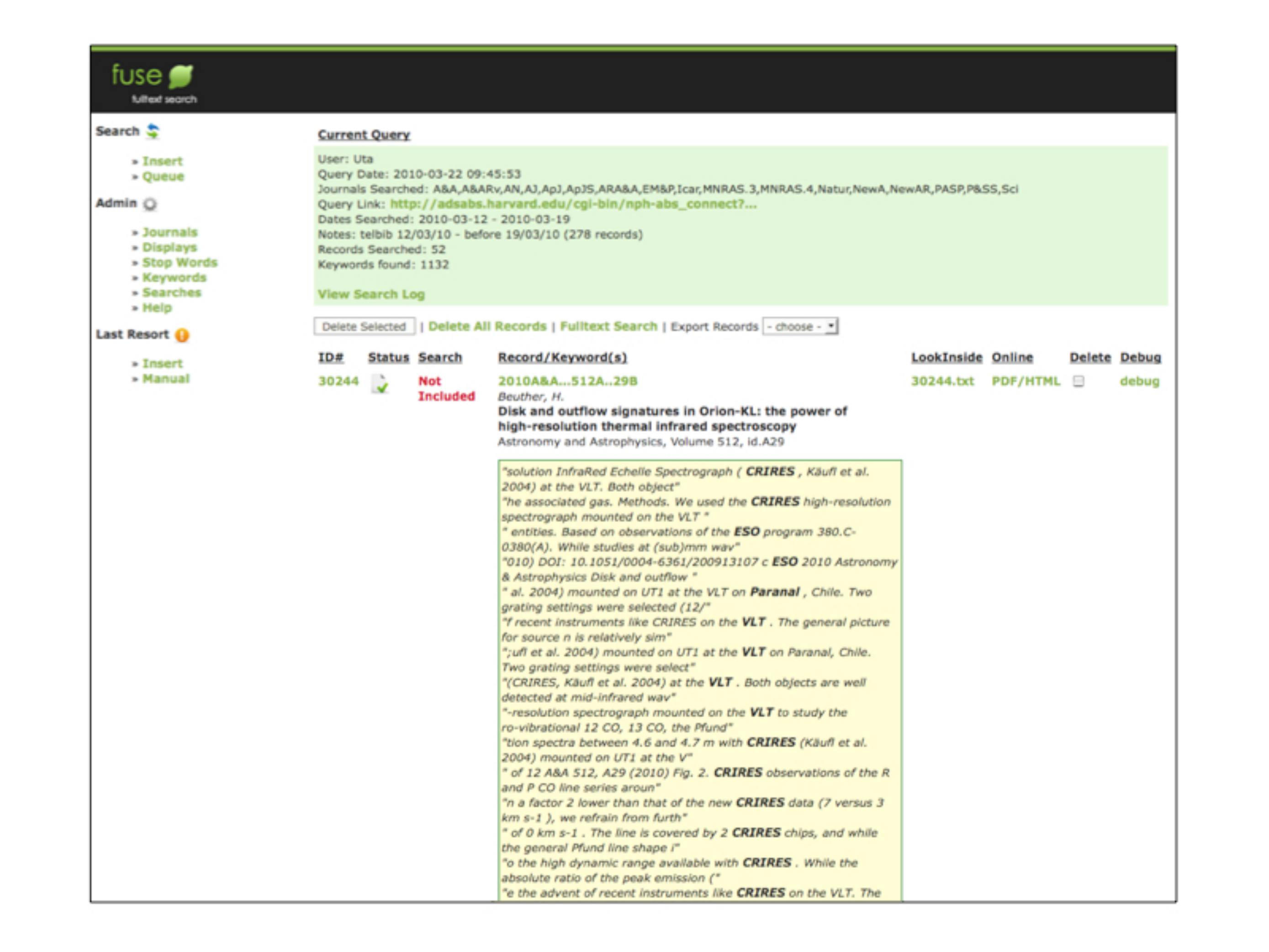}}
\caption{Screenshot of the ESO Full-Text Search tool (FUSE).}
\end{figure}

It is important to note that FUSE can only be used as a help application in order to spot potential 
candidates for inclusion in telescope bibliographies. By no means does it replace the human (intellectual) 
investigation that is necessary in order to determine whether an observatory's selection criteria for 
inclusion in the bibliography are actually met by the paper.
  
Another important tool is telbib, a content management software also developed by the ESO Library. It  is 
used to establish and maintain a database of records with bibliographic information about papers pertaining 
to a telescope bibliography, to store additional metadata (e.g., tags to describe observing facilities and 
programs), and to generate reports and statistics. The librarians' user interface provides access to records 
through a large variety of search criteria (Fig. 3). A public user interface is also available, even though 
with less detailed query features\footnote{\tt http://www.eso.org/libraries/telbib.html}. 
Using telbib, bibliography compilers can import bibliographic information 
and other metadata from the ADS, including author affiliations and number of citations. Observatory-defined 
tags and keywords can then be added, as well as specific information about observing programs, modes 
(e.g., visitor or service mode) and types (normal, large, guaranteed time observing types, etc.) .

\begin{figure}
\centerline{\includegraphics[height=10cm,angle=0]{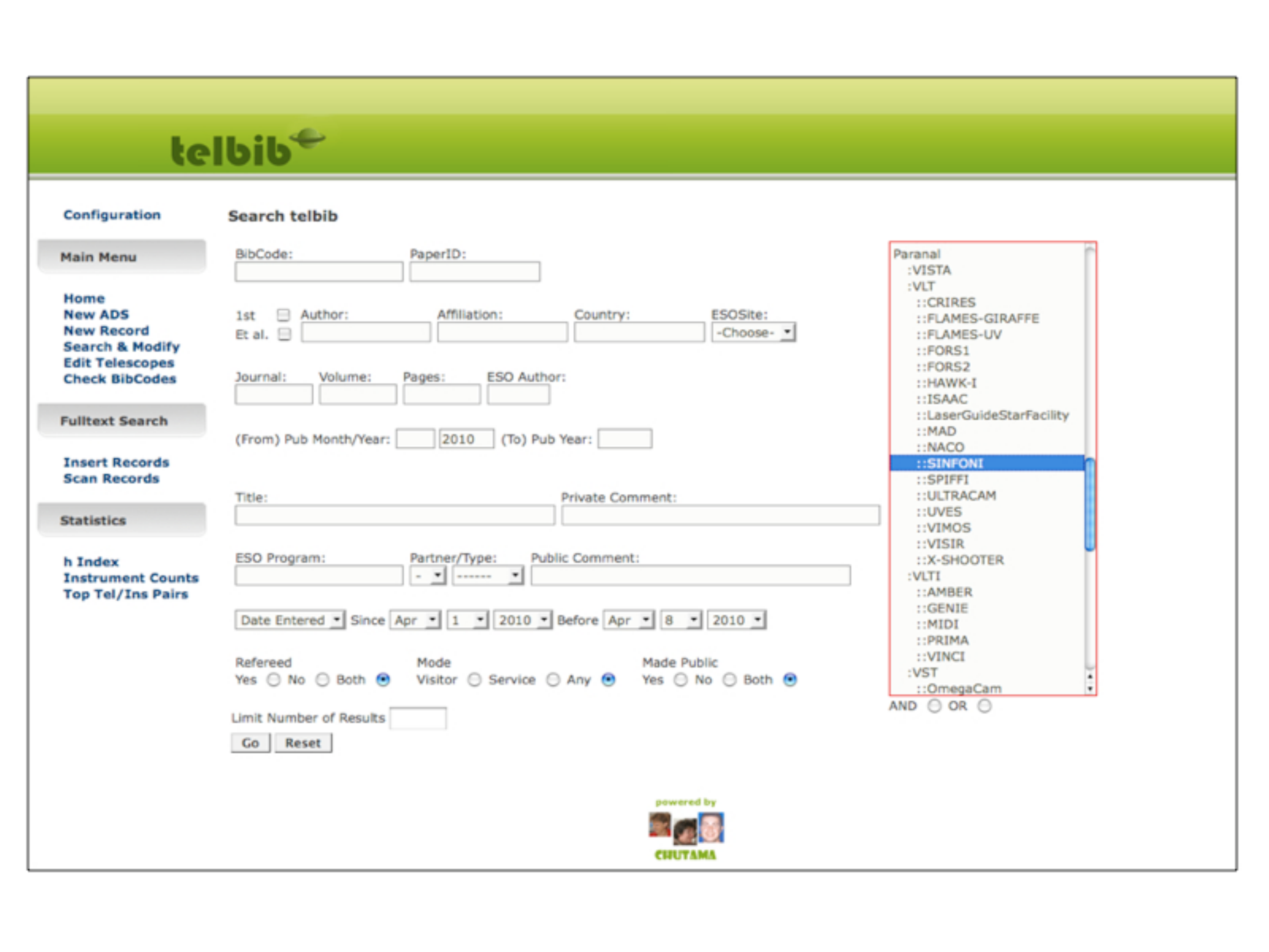}}
\caption{Screenshot of the ESO telbib tool.}
\end{figure}

Telbib is also used to generate reports and statistics either on demand or regularly, for instance the 
``Basic ESO Statistics" document\footnote{\tt http://www.eso.org/sci/libraries/edocs/ESO/ESOstats.pdf}
that is available from the Libraries' home page\footnote{\tt http://www.eso.org/libraries/}. It provides 
information on ESO publication and citation statistics at large as well as for specific instruments, 
identifies the ``ESO Top 20" papers, and looks at ESO publications in comparison with other large 
observatories. 

A more detailed description of FUSE and telbib including their development and features can be found in 
Erdmann \& Grothkopf (2010).

FUSE is currently in use at ESO, STScI, Gemini, Subaru, and the Carnegie Observatories. The libraries of 
SAAO (South African Astronomical Observatory) in Cape Town, South Africa, and IUCAA (Inter-University 
Centre for Astronomy and Astrophysics) in Pune, India, are in the process of installing the software. 
Because of this obvious duplication of efforts by conducting searches locally, recently the idea of 
creating a central search facility at the ADS has been discussed. Compilers of bibliographies could then 
run tailored searches at the ADS without the need to download PDFs locally for inspection. The search 
feature would only highlight potentially relevant matches by showing brief excerpts of text; access to 
the full-texts would still be governed by the publishers through subscriptions. 

\section{Beyond citations}

As already mentioned, studying citations to measure scholarly output is problematic, despite their 
wide-spread use for this purpose. Lists of citations are often incomplete, references are incorrect, 
various abbreviations for the same journal are used in parallel and are sometimes not attributed to 
the correct publication, and the typical citing behavior often favors well-known researchers instead 
of referencing younger, unknown authors. 

Are there other, possibly better ways of evaluating scientific impact? Compilers of telescope bibliographies 
are experimenting with other measures. Some of them are described below. 

It is well known that the ADS provides information about citations for each paper in their database; these 
are limited to those citing papers that also reside in the ADS database, hence they are not entirely 
complete. However, in the area of astronomy this is a minor disadvantage as the ADS coverage of papers 
is extremely large. 

Few users are aware that ADS also gathers information about so-called Reads. Reads are defined as those 
instances where users access more detailed information than is available in the standard brief results 
list displayed after an ADS search. Hence, a Read can mean that a user looked at the abstract, the list 
of references, or the citing papers. Also any time the electronic version of an article is accessed, be 
it in PDF or in HTML format, counts as a Read. 

The number of Reads typically differs considerably from the number of citations an article has gathered. 
More importantly, the distribution of Reads versus citations as a function of years reveals that many 
papers which are not cited frequently anymore still are used regularly by the community. 

An even more telling statistic about actual usage could be obtained by limiting Reads to the actual request 
of full-texts, i.e., PDFs and HTML downloads. Such numbers are traced by the ADS, but in contrast to Reads, 
they cannot be retrieved through the ADS user interface. However, the ADS team is always helpful in 
providing such statistics on request. 

Another project that operates along the same line of thought 
is Citebase\footnote{\tt http://www.citebase.org/}. Citebase is a citation index 
that ``harvests pre- and post-prints (mostly author self-archived)" from arXiv and other repositories and 
lists them together with papers that cite them. In addition, they provide information about downloads by 
country, by date and by organization. Unfortunately, they are still in experimental stage, and users are 
cautioned not to use the results for academic evaluation. 

A newcomer in the area of evaluating research impact is the idea to look at social networking platforms and 
forum discussions. In astronomy CosmoCoffee\footnote{\tt http://www.cosmocoffee.info/}
 might be one such discussion group that reveals the community's 
interest in specific papers. Registered users refer to manuscripts posted on arXiv and suggest discussion 
among their peers. In order to track impact, bibliography compilers could track which arXiv preprints are 
discussed so that these papers can be linked to the final versions once they are published, and the number 
of comments and discussion threads specific papers generate can be traced. 

\section{Ongoing projects}

In order to provide even more informative reports and to anticipate future interests of management and 
funding organizations, it is desirable to import further metadata from the ADS, for instance subject terms 
and keywords (to gain knowledge about the specific subject areas in which users of observational data 
publish), full information of citing papers (in order to eliminate self-citations), all other available 
links (e.g., Digital Object Identifiers (DOI), eprint IDs to establish links between eprints and published 
papers), as well as the author gender. The latter might be of interest to researchers involved in topics 
such as ``Women in Astronomy"; identifying an author's gender may be feasible in future with the help of 
initiatives like ORCID (Open Researcher and Contributor ID)\footnote{\tt http://www.orcid.org/}  that are 
hoped to solve the name ambiguity problem in scholarly research.  

It is also intended to establish additional links, for instance between telescope bibliography records 
and press releases that feature specific papers, as well as record detailed information about the observing 
dates so that the delay from data acquisition to the publication can be computed. 

\section{Conclusion}

The task of compiling telescope bibliographies has evolved considerably during recent years, providing ever 
more detailed and sophisticated information about the papers contained in the databases of large 
observatories. This trend will continue in the future as management and funding agencies rely increasingly 
on bibliometric studies in order to evaluate research output. It is important that compilers collaborate and 
exchange ideas and solutions to common problems in order to develop best practices and recommendations for 
establishing, maintaining, and using telescope bibliographies in a standardized, reproducible manner. 

\section*{Acknowledgments}

The ESO Telescope Bibliography (telbib) and Full-Text Search tool (FUSE) use NASA's Astrophysics Data 
System (ADS) Abstract Services. The authors wish to thank the ADS team for their excellent service to 
the entire community.


\begin{thebibliography}{99}
\bibitem{}
Erdmann, C. \& Grothkopf, U., 2010: Next generation bibliometrics and the evolution of the ESO Telescope 
Bibliography, in: {\em Library and Information Services in Astronomy VI}, Isaksson, E., Lagerstrom, J., 
Holl, A. and Bawdekar, N.  (eds.), Astronomical Society of the Pacific, San Francisco, ASP Conference Series
{\bf 433}, p. 81
\bibitem{}
Hirsch, J.E., 2005: An index to quantify an individual's scientific research output, {\em Proc. Nat. Acad. 
Sci.}, {\bf 102}, 16569, DOI 10.1073/pnas.0507655102
\bibitem{}
Lagerstrom, J, 2010: Comparison of methods for creating telescope bibliographies, in: {\em Library and 
Information Services in Astronomy VI}, Isaksson, E., Lagerstrom, J., Holl, A.  and Bawdekar, N. (eds.), 
Astronomical Society of the Pacific, San Francisco, ASP Conference Series {\bf 433}, p. 89
\end{thebibliography}
\end{document}